\documentclass[10pt, a4]{article}

\usepackage[english]{babel}
\usepackage[T1]{fontenc}
\usepackage[utf8]{inputenc}
\usepackage{hyperref}
\usepackage{color}

\usepackage{booktabs}
\usepackage{graphicx}

\usepackage{amsthm}
\usepackage{amsmath}
\usepackage{amssymb}

\newtheorem{theorem}{Théorème}
\newtheorem{definition}{Definition}
\newtheorem{exemple}{Exemple}

\begin{document}

\title{Techniques d'anonymisation tabulaire : concepts et mise en \oe uvre}
\author{Benjamin Nguyen$^1$ \and Claude Castelluccia$^2$}
\date{%
\small{$^1$INSA Centre Val de Loire, Laboratoire d'Informatique
Fondamentale d'Orléans \\
$^2$Inria  \\   }%
    \today
}
\maketitle

\section*{Résumé}

Dans ce document, nous présentons l'état de l'art des techniques d'anonymisation
pour des bases de données classiques (i.e. des tables), à destination d'un
public technique ayant une formation universitaire de base en mathématiques et
informatique, mais non spécialiste. L'objectif de ce document est d'expliquer
les concepts permettant de réaliser une anonymisation de données tabulaires,
et de calculer les risques de réidentification. Le document est largement
composé d'exemples permettant au lecteur de comprendre comment mettre en \oe
uvre les calculs.

\newpage

\tableofcontents

\newpage

\section*{Introduction : le secret statistique}

En France, il existe depuis 1951 la loi sur \emph{l'obligation, la coordination,
et le secret en matière de statistiques}~\cite{secretStat}.
L'histoire permet aisément de comprendre pourquoi ces questions ont vu le jour à
la sortie de la seconde guerre mondiale. L'idée fondamentale défendue dans cette
loi est que ``\emph{les renseignements individuels (\ldots) et ayant trait à la
vie personnelle et familiale et, d'une manière générale, aux faits et
comportements d'ordre privé, ne peuvent faire l'objet d'aucune communication}''.
De même en ce qui concerne l'Europe, ce principe est affirmé par l'article 338
du traité de l'Union~\cite{secretStatEU} : ``\emph{L'établissement des
statistiques se fait dans le respect de l'impartialité, de la fiabilité, de
l'objectivité, de l'indépendance scientifique, de l'efficacité au regard du coût
et de la confidentialité des informations statistiques;}''. L'idée fondamentale
qui existe depuis près de 70 ans est donc qu'il faut toujours s'assurer de la
confidentialité des données privées lors d'opérations de traitement de
données. L'INSEE a publié un guide du secret statistique~\cite{secretStatINSEE},
précisant que ``\emph{pour les tableaux fournissant des données agrégées sur les
personnes physiques, le secret statistique impose qu’on ne puisse pas connaître
ou déduire des informations les concernant}''. Il est intéressant de noter que
ce guide n'explique pas comment atteindre le secret statistique, mais montre des
cas simples où le secret statistique n'est pas réalisé : ``\emph{Par exemple, si
un tableau donne pour une commune la répartition par âge et situation
matrimoniale, et que toutes les personnes âgées de 50 à 59 ans ont toutes pour
état matrimonial « divorcé », le secret statistique n’est plus respecté dans ce
tableau. En effet, si l’on sait par ailleurs que quelqu’un a entre 50 et 59 ans,
le tableau nous informe que cette personne est divorcée}''. Nous verrons que ce
problème persiste pour certaines techniques d'anonymisation.

\section{Anonymisation : les principes}

\subsection{L'anonymisation dans la législation}

Depuis 1978 et la loi ``Informatique et libertés''~\cite{infoLib}, le principe
de protection affirmé par la loi a évolué. En effet, les lois relatives au
traitement de données personnelles précisent ce qu'est une telle donnée, ce qui
permet de savoir par contraposée ce qu'est une donnée anonymisée :
``\emph{Constitue une donnée à caractère personnel toute information relative à une personne
physique identifiée ou qui peut être identifiée, directement ou indirectement,
par référence à un numéro d'identification ou à un ou plusieurs éléments qui lui
sont propres. Pour déterminer si une personne est identifiable, \textbf{il
convient de considérer l'ensemble des moyens en vue de permettre son identification dont
dispose ou auxquels peut avoir accès le responsable du traitement ou toute autre
personne}}.'' La législation française était donc à l'origine très rigide
concernant une donnée anonyme, puisque la définition avait trait à une
obligation de résultat (impossibilité que quiconque puisse remonter à la donnée
originale). Nous verrons qu'une telle contrainte n'est pas compatible avec les
méthodes d'anonymisation proposées, qui conservent toujours un risque de
réidentificationn. Par conséquent avec cette définition, toute donnée devait
être considée comme une données personnelle, ce qui n'est clairement par
l'objectif recherché. En effet, nous allons voir qu'il est possible de
quantifier la probabilité du risque de réidentification, et on pourra ainsi
estimer qu'une donnée sera anonyme si ce risque est acceptable.

\begin{exemple}[Données anonymes pré-RGPD]
Un enregistrement, même utilisant un algorithme de chiffrement réputé sûr comme
AES, ayant produit le n-uplet suivant (Nom :
``sEujSEsWzHioxae70aKE6w=='', Age : ``27'', Salaire :
``36800K'', Département : ``75'') \emph{n'est pas anonyme} dans la mesure où la
personne ayant chiffré le n-uplet est capable de le déchiffrer, puisqu'elle connait la clé de chiffrement
utilisée\footnote{Pour ceux qui s'y intéressent, la clé de chiffrement utilisée
ici est une clé basique : ``ZZZZZZZZZZZZZZZZ''.}.
\end{exemple}

Depuis mai 2018 et l'entrée en vigueur du Règlement Général sur la Protection
des Données~\cite{rgpd} (RGPD ou GDPR en anglais), cette obligation de résultat
s'est transformée en obligation de moyens, indiquée dans le
considérant 26, éclairant l'interprétation de la signification de ce qu'est une donnée
identifiante :
``\emph{Il y a lieu d'appliquer les  principes relatifs à  la  protection  des  données à  toute  information
concernant  une personne physique identifiée ou identifiable. (\ldots) Pour
déterminer si une personne physique est identifiable, il convient de prendre en considération
l'ensemble des moyens raisonnablement susceptibles d'être utilisés par le
responsable du traitement ou par toute autre personne pour identifier la
personne physique directement ou indirectement, tels que le ciblage. Pour
établir si des moyens sont raisonnablement susceptibles d'être utilisés pour
identifier une personne physique, il convient de prendre en considération
l'ensemble des facteurs objectifs, tels que le coût de l'identification et le
temps nécessaire à celle-ci, en tenant compte des technologies disponibles au
moment du traitement et de l'évolution de celles-ci.}''  La définition d'une
donnée identifiante dans la loi française est désormais celle référencée dans
l'article 4 du RGPD. Il semble donc raisonnable de considérer la définition
\emph{relaxée} à l'obligation de moyens. Faute de cette
interprétation, chercher à anonymiser des données tout en conservant une certaine utilité ne
fait pas sens. Il faut donc accepter une part de risque de réidentification
lorsqu'un processus d'anonymisation a lieu, et chercher à quantifier la
probabilité de désanonymisation, l'impact, mais également le bénéfice qui pourra
être obtenu suite au traitement des données. La définition de l'anonymisation
se fait donc par rapport à la question de la réidentification.

\begin{exemple}[Données anonymes post-RGPD]
Considérons la table~\ref{t1}, générée en anonymisant une partie des
données\footnote{Ces salaires sont disponibles en accès libre dû au règlement sur le ``fair play financier'' de l'EUFA. On pourrait
s'interroger sur l'impact de ce règlement sur le droit à la vie privée des
joueurs. Les données ont été extraites de :
\href{https://www.linternaute.com/sport/foot/1433459-salaires-de-ligue-1-le-nom-du-joueur-le-mieux-paye-revele-vendredi/}{https://www.linternaute.com/sport/foot/1433459-salaires-de-ligue-1-le-nom-du-joueur-le-mieux-paye-revele-vendredi/}
} de la table~\ref{tori} (on a gardé uniquement les joueurs du PSG).
\begin{table}

\begin{center}
\begin{tabular}{|l|l|l|l|}
\hline
\textbf{ID}& \textbf{age} & \textbf{Club} & \textbf{Salaire}\\
\hline
\tiny{Thiago Silva} & 35 & PSG & 1160K\\
\tiny{Edison Cavani} & 32 & PSG & 1500K\\
\tiny{Kylian Mbappé} & 20 & PSG & 1730K\\
\tiny{Neymar Jr.} & 27 & PSG  & 3060K\\
\tiny{Dimitri Payet} & 32 & OM & 500K\\
\tiny{Luiz Gustavo} & 32 & OM  & 500K\\
\hline
\end{tabular}
\end{center}
\caption{Salaires 2019\label{tori}}

\begin{center}
\begin{tabular}{|l|l|l|l|}
\hline
\textbf{ID}& \textbf{age} & \textbf{Club} & \textbf{Salaire}\\
\hline
\tiny{dH7SdankcIhHDE1ATvErkg} & [30;39] & PSG & 1160K\\
\tiny{fTRVz9bY2mHguqsmuPHtvw} & [30;39] & PSG & 1500K\\
\tiny{x4TUcj1FQZkfSfnIeL05NA} & [20;29] & PSG & 1730K\\
\tiny{jtILvRsZLVUETwlExAzpvw} & [20;29] & PSG  & 3060K\\
\hline
\end{tabular}
\end{center}

\caption{Salaires 2019 du PSG, table anonymisée\label{t1}}

\end{table}
La première ligne est une valeur hachée du nom utilisant une clé de chiffrement
qui a été détruite (effacée). Dans ce cas, même si un attaquant connait l'age de
la personne et son club, elle ne sera pas capable de déduire son salaire avec
une probabilité supérieure à 50\%. Par exemple, Cavani, agé de 32 ans et jouant
au PSG peut avoir un salaire de 1.160 millions ou 1.500 millions.
\end{exemple}

\subsection{Objectif d'un processus d'anonymisation}

Nous notons $\mathcal{D}$ une base de données, et $\mathcal{A}_X(\mathcal{D})$
sa version anonymisée selon un mécanisme $\mathcal{A}_X$.

\begin{definition}[Réidentification]
La réidentification est un processus (ou algorithme) prenant en entrée un jeu de
données (anonymes), des connaissances annexes et cherchant à apparier des
données anonymes avec des individus du monde réel.
\end{definition}

\begin{exemple}[Réidentification]
Considérons de nouveau la Table~\ref{t1}. Supposons que la connaissance annexe
d'un attaquant soit de savoir qu'il n'y a que 3 joueurs du PSG qui gagnent plus
de 1.5 millions d'euros par mois (Neymar Jr., Mbappé et Cavani), et que l'age
de Cavani est de 32 ans, alors il est possible à cet attaquant de déduire le
salaire de Cavani de la Table~\ref{t1} (1.5 millions).
\end{exemple}

Un processus d'anonymisation doit idéalement protéger contre toute
réidentification, et doit \emph{a minima} quantifier le risque de
réidentification de la base de données. Cette réidentification peut être
quantifiée par rapport à différents modèles~\cite{DBLP:journals/jamia/EmamD08} :
le modèle du journaliste, du procureur, et du marketeur. Le modèle du
journaliste considère que l'attaque est réussie si l'attaquant (le journaliste)
arrive à désanonymiser n'importe quelle personne du jeu de données, et à la
retrouver dans le monde réel. Le modèle du procureur considère que l'individu
auquel on s'intéresse est dans la base de données, qu'on connait toutes les
informations sur lui, et que l'attaque est réussie si on réussi à le
désanonymiser. Le modèle du marketeur cherche à apparier un maximum
d'enregistrements pour lesquels on ne connais que le quasi-identifiant (voir plus bas).

\begin{exemple}[Risque du journaliste]
Toujours dans l'exemple de la Table~\ref{t1}, le modèle du journaliste fera
comme hypothèse que le journaliste connait toutes les données publiques
relatives aux individus présents dans la base. Il connaitra donc l'age et
l'équipe des individus : (Mbappé, 20, PSG), (Neymar Jr., 27, PSG), (Cavani,
32, PSG), (Silva, 35, PSG). Ici, il est incapable de différencier Cavani de
Silva et Mbappé de Neymar Jr. soit une chance sur deux à chaque fois. Le risque de retrouver au moins l'un
des quatre individus est donc égal à :
$R_{J}=1-(\frac{1}{2})^2=0.75$
\end{exemple}

\begin{exemple}[Risque du procureur]
Le modèle du procureur fera
comme hypothèse que le procureur connait toutes les données publiques
relatives à l'individu qui l'intéresse présent dans la base. Il
connaitra donc l'age et l'équipe de : (Mbappé, 20, PSG). Ici, il est incapable
de différencier Mbappé de Neymar Jr. Le risque de retrouver le salaire de Mbappé est donc égal à :
$R_{P}('Mbappe')=\frac{1}{2}=0.5$ On voit que le risque du procureur dépend de
l'individu qu'il cherche à réidentifier. On peut donc définir le risque maximal du
procureur comme étant $R_{P}=\max_{i\in\mathcal{D}}(R_{P}(i))$
\end{exemple}

\begin{exemple}[Risque du marketeur]
Le modèle du marketeur fait la même hypothèse que celui du journaliste,
toutefois il cherche à calculer le nombre d'appariements réussis. On peut le
voir comme une espérance normalisée du nombre de réidentifications réussies.
$R_{M}=\Sigma_{i\in\mathcal{D}} p(i)/|\mathcal{D|}=0.5$

\end{exemple}

\begin{definition}[Utilité]
L'utilité $U(r(\mathcal{D}))$ d'une requête $r(\mathcal{D})$ d'analyse de
données se mesure par une métrique $\mathtt{M}$ qui indique la différence entre
la valeur de $U(r(\mathcal{D}))$ et la valeur de
$U(r'(\mathcal{A}_X(\mathcal{D})))$, où $r'$ peut être $r$ ou une version
modifiée de $r$ pour prendre en compte le processus d'anonymisation.
\end{definition}

\begin{exemple}[Utilité du calcul du salaire par age]
Supposons que nous souhaitons calculer le salaire par age sur la base de données
constituée des 4 joueurs de notre exemple. Le résultat est indiqué dans la
Table~\ref{t2}.

\begin{table}
\begin{center}
\begin{tabular}{|l|l|l|l|}
\hline
\textbf{age} & \textbf{Salaire}\\
\hline
35 & 1160K\\
32 & 1500K\\
20 & 1730K\\
27 & 3060K\\
\hline
\end{tabular}
\end{center}

\caption{Résultats exacts : $Q1$\label{t2}}
\end{table}

Si nous lançons la même requête sur les données anonymisées nous obtiendrons le
résultat présenté dans la Table~\ref{t3}. Nous pouvons évaluer l'utilité du
deuxième calcul grâce à la fonction d'utilité suivante, qui calcule la moyenne
de l'erreur normalisée (toute autre fonction d'évaluation d'erreur pourrait
être pertinente).

$ \mathtt{M}(U(Q_1(\mathcal{D}),
U(Q'_1(\mathcal{A}_X(\mathcal{D}))))=1/4\times(|\frac{1330-1160}{1160}|+|\frac{1330-1500}{1500}|+|\frac{2395-1730}{1730}|+|\frac{2395-3060}{3060}|)=0.215$

\begin{table}
\begin{center}
\begin{tabular}{|l|l|l|l|}
\hline
\textbf{age} & \textbf{Salaire}\\
\hline
[30;39] & 1330K \\
    
[20;29] & 2395K \\
\hline
\end{tabular}
\end{center}

\caption{Résultats sur données anonymes : $Q'_1$\label{t3}}
\end{table}

\end{exemple}

Il est entendu que l'objectif de la publication d'une base de données est
d'effectuer des analyses de données la concernant. Il s'agit donc pour le
responsable de l'anonymisation de proposer un \emph{compromis} acceptable (et
le meilleur possible) entre d'une part la sécurité (anonymat) de la base de
données, c'est-à-dire la difficulté de la réidentification, et d'autre part
l'utilité (estimée ou calculée) de la base anonymisée.

\subsection{Approche Pratique}

{\bf Du cas par cas.} Il n'existe malheureusement pas de solution
d'anonymisation universelle qui s'appliquerait à tous les types d'applications
et de données.  Une solution d'anonymisation est souvent le  resultat d'un
(long) travail d'optimisation entre les garanties en terme de sécurité qu'on
souhaite fournir et l'utilité des données.
Une solution d’anonymisation doit donc être développée au cas par cas et adaptée
aux usages prévus et aux données traitées.
Par exemple, des données sensibles, comme des données de santé, nécessiteront
probablement des solutions d'anonymisation plus robustes que des données moins
sensibles, comme par exemple des données de mobilité dans un musée. Par
ailleurs, le type de publication envisagée est aussi important à considerer: des
données publiées sur le Web, en libre accès, necessiteront des garanties
plus fortes que des données partagées avec un partenaire industriel avec qui un
contrat juridique pourra être éventuellement signé.

{\bf Anonymisation et analyse de risques.} Il est donc souvent souhaitable de
combiner le travail de conception de solution d'anonymisation avec une analyse
de risques.
Cette tâche consiste à décrire précisement la nature des données utilisées, les
traitements mis en œuvre, les différents acteurs, en considérant tant les
aspects techniques qu'opérationnels. Il conviendra ensuite d'evaluer les risques
sur la sécurité des données (confidentialité, intégrité et disponibilité) ainsi
que  leurs impacts potentiels sur la vie privée. Ce travail d'analyse permettra
de développer la solution la mieux adaptée aux besoins et aux contraintes
existants \cite{Art29op}.

{\bf Evaluation} Afin d'évaluer les solutions d’anonymisation, le G29 (le
groupement des autorités de protection des données européennes)  propose trois
critères : l'individualisation, la corrélation, l'inférence \cite{Art29op}.
Ainsi pour ``démontrer'' qu'une solution est correcte et conforme au RGPD, il faut
demontrer que les données anonymisées ne permettent  plus d'isoler les données
qui appartiennent à un individu, de relier entre eux des ensembles de données
distincts concernant un même individu  et de déduire de l’information sur un
individu. Dans le cas où un des trois critères n'est pas respecté, le G29
stipule que les données ne pourront être considérées comme anonymisées
uniquement si une analyse détaillée  démontre que les risques de 
ré-identification sont maîtrisés! Par ailleurs, étant donné que les techniques
de ré-identification s'améliorent, il est indispensable de ré-évaluer
régulièrement le caractère anonyme des données produites.

\subsection{La pseudonymisation}

Une erreur courante consiste à considérer la pseudonymisation comme une solution
d'anonymisation.
La pseudonymisation est une technique simple qui consiste à remplacer la valeur
d'un attribut ``identifiant'' (par exemple un nom) par une autre valeur, un
``pseudo'' (comme le cas de l'Exemple 1). Ce ``pseudo'' peut être généré
indépendamment de la valeur d'origine et en etre dérivé (par exemple, en
appliquant une fonction de hachage). Clairement, le pseudonymisation permet de
reduire le risque de mise en corrélation d'un ensemble de données avec
l'identité originale d'un sujet,  mais une ré-identification indirecte est
possible en utilisant, par exemple, les autres attributs \cite{Nature13}. Par
conséquent, il est important de rappeler que la pseudonymisation est une mesure
de sécurité utile, et qu'il faut encourager, mais ne constitue pas une solution
d'anonymisation à part entière.

\section{Techniques d'anonymisation}

\subsection{L'anonymisation moderne : naissance du $k$-anonymat}

On peut dater la problématique moderne de l'anonymisation de la fin des années
1990 avec la publication par Sweeney de plusieurs articles où elle propose le
concept de $k$-anonymat ($k$-anonymity),
dont nous citons le plus connu~\cite{DBLP:journals/ijufks/Sweene02a} publié en
2002. En effet, à l'époque lorsqu'on parlait d'anonymisation, on faisait
référence au concept désormais connu sous le nom de \emph{pseudonymisation}.
Comme indiqué précédemment, la pseudonymisation est le remplacement de toutes
les données directement identifiantes (comme le numéro de sécurité sociale) par une valeur aléatoire
(pseudonyme). Sweeney a montré dans~\cite{DBLP:journals/ijufks/Sweene02a} qu'il
était possible de ré-identifier une base de données relationnelle pseudonymisée
(SQL ou plus généralement un fichier tabulaire) en utilisant ses
\emph{quasi-identifiants}.

\begin{definition}[Quasi-identifiant]
Soit $T(A_1,\ldots, A_n)$ une table composée de $n$ attributs $A_i$. On appelle
quasi-identifiant $Q$ de T un ensemble d'attributs $Q=\{A_i,\ldots, A_j\}
\subseteq \{A_1,\ldots,A_n\}$ dont la publication doit être contrôlée de manière
suivante : Pour être un quasi-identifiant, $Q$ doit être tel que la requête
\texttt{
SELECT $A_1,...,A_n$ \\
FROM T \\
GROUP BY $A_1, ... , A_n$ \\
HAVING COUNT = 1 \\
}
retourne un résultat non vide. 
\end{definition}
On pourra noter, en terme de conception de base de données, que si $Q$ est une
clé alors $Q$ est un QID mais pas forcément l'inverse. Si $Q$ était une clé
alors la requête précédente donnerait très exactement la projection de $T$ sur
$Q$.

On peut décider, indépendamment de toute connaissance annexe, que $Q$ est un
QID, puisqu'il s'agit d'une propriété de l'instance de la table. Par contre, le
risque de réidentification est différent et dépend de la connaissance annexe de
l'attaquant : si on identifie un QID mais qu'on est \emph{sûrs} que l'attaquant
ne possède aucune base avec ce même QID on pourra estimer que le risque de
réidentification via ce QID est faible. A l'inverse, identifier un QID sur une
table contenant des données sensibles, ainsi qu'une base de données annexe
contenant ce même QID ainsi que des données identifiantes présente un risque
réel.

\begin{exemple}[QID des footballeurs]
On peut considérer que le couple (age, club) forme un quasi-identifiant. Ainsi,
Thiago Silva est le seul joueur du PSG à être agé de 35 ans. Mbappé est le seul
joueur du PSG à être agé de 20 ans. Toutefois il y a plusieurs joueurs agés de
27 ans (Neymar, Kurzawa, Sarabia).
\end{exemple}

Rappellons que la logique de la définition repose sur le fait qu'il est possible
de retrouver dans une \emph{autre} base de données à la fois $Q$ mais aussi un
attribut (ou un ensemble d'attributs comme le couple nom/prénom) identifiant.
Notons que pour deux quasi-identifiants $Q_1 \subseteq Q_2$ on a $Q_1$ est un
quasi-identifiant $\rightarrow Q_2$ est un quasi-identifiant. Lors de l'étude
des quasi-identifiants, il convient donc de s'intéresser au(x)
quasi-identifiant(s) maximal(aux) par rapport à l'inclusion.

Sweeney définit alors le critère de $k$-anonymat, et propose une garantie de
sécurité par rapport à la ré-identification.

\begin{definition}[$k$-anonymat]
La publication d'une version anonyme $P_\mathcal{D}$ d'une base de données
$\mathcal{D}$ respecte le critère de $k$-anonymat par rapport à un
quasi-identifiant $Q$ ssi chaque valeur de $q \in Q$ dans
$P_\mathcal{D}$ apparaît au moins $k$ fois. On parle alors de \emph{classe
d'équivalence} pour tous les n-uplets qui ont la même valeur $q$.
\end{definition}

\begin{exemple}[$2$-anonymat des footballeurs]
La Table~\ref{t4} est $2$-anonyme par rapport au QID (\textbf{age},
\textbf{club}).
On voit toutefois qu'on a été obligé de modifier le domaine de définition de
l'attribut \textbf{age} de l'année à la décennie.
\begin{table}
\begin{center}
\begin{tabular}{|l|l|l|l|}
\hline
\textbf{age} & \textbf{Club} & \textbf{Salaire}\\
\hline
[30;39] & PSG & 1160K\\

[30;39] & PSG & 1500K\\

[20;29] & PSG & 1730K\\

[20;29] & PSG  & 3060K\\

\hline
\end{tabular}
\end{center}

\caption{Données $2$-anonymes\label{t4}}
\end{table}
\end{exemple}

Le $k$-anonymat peut être vu comme une \emph{contrainte} que doit respecter une
version publiée du jeu de données. Plusieurs algorithmes permettant de respecter
une telle contrainte existent~: des algorithmes basés sur de la suppression
(effacement de la valeur d'un attribut ou effacement d'un n-uplet), des
algorithmes basés sur de la généralisation (modification de la valeur d'un
attribut pour le généraliser, tout en conservant la même signification, par
exemple généralisation de la commune vers le département ou la région), ou des
algorithmes combinant ces deux techniques, comme le préconise Sweeney
dans~\cite{DBLP:journals/ijufks/Sweene02a}. Une fois qu'on a procédé à cette
transformation pour respecter la contrainte sur les quasi-identifiants, on peut
publier les données, en leur associant les autres informations (considérées
comme des données sensibles).

{\bf Optimalité des algorithmes} L'une des questions débattues au sein de la
communauté était celle de \emph{l'optimalité} d'une telle anonymisation, c'est-à-dire une
transformation qui ferait perdre le moins d'informations, tout en respectant la
contrainte~\cite{Meyerson:2004:COK:1055558.1055591}. Meyerson et Williams ont
montré que trouver la valeur optimale est difficile
(NP-difficile), même s'il existe des $O(k \log k)$-approximations calculables en temps polynomial.

{\bf Risque de réidentification} La protection \emph{revendiquée} par une base
de données anonymisée selon la contrainte de $k$-anonymat est que chaque n-uplet
étant confondu avec $k-1$ autres, la probabilité de retrouver le n-uplet
correct si on connait les valeurs exactes du quasi-identifiant est de $1/k$. Il est
important de souligner que la garantie proposée, comme toutes les garanties en
anonymisation, est une garantie \emph{probabiliste}. Il convient donc au
responsable de traitement mettant en oeuvre l'anonymisation de décider du risque
de réidentification qu'il est prêt à accepter, et de choisir le paramètre de $k$
en conséquence.

\subsection{Faiblesse du modèle du $k$-anonymat}

L'intérêt principal du $k$-anonymat est qu'il est facile à comprendre. Les
algorithmes permettant de l'implémenter sont également assez rapides (il ne faut
guère plus de quelques secondes pour anonymiser une base de données de plusieurs
dizaines ou centaines de milliers de lignes). Toutefois, le modèle n'est pas
robuste par rapport aux \emph{attaques d'homogénéité}, comme proposées par
Machanavajjhala \emph{et al.}~\cite{DBLP:journals/tkdd/MachanavajjhalaKGV07}.
Une telle attaque a lieu lorsque les valeurs sensibles associées à une
valeur donnée de quasi-identifiant sont toutes identiques. Dans ce cas, on peut
déduire que toutes les personnes ayant cette valeur de quasi-identifiant ont la
même donnée sensible, qu'on est capable de déduire.

Il est donc impossible de donner \emph{a priori} une garantie sur le risque de
réidentification, ce qui fait qu'une utilisation du $k$-anonymat seul ne
présente pas de garanties d'anonymat raisonnables.

\begin{exemple}[Les footballeurs marseillais]
Considérons la base de données contenant des joueurs d'autres clubs, construite
à partir de l'intégralité de la Table~\ref{tori}, et présentée de manière
2-anonyme dans la Table~\ref{t5}. Si on sait que seuls les joueurs Payet et
Gustavo ont 32 ans et jouent à Marseille, on sera capable de déduire que leur
salaire est de 500, puisqu'ils ont tous les deux le même salaire.

\begin{table}
\begin{center}
\begin{tabular}{|l|l|l|l|}
\hline
\textbf{age} & \textbf{Club} & \textbf{Salaire}\\
\hline
[30;39] & PSG & 1160K\\

[30;39] & PSG & 1500K\\

[20;29] & PSG & 1730K\\

[20;29] & PSG  & 3060K\\

[32] & OM  & 500K\\

[32] & OM  & 500K\\

\hline
\end{tabular}
\end{center}

\caption{Données $2$-anonymes avec un risque de ré-identification\label{t5}}
\end{table}
\end{exemple}

\subsection{Extensions du modèle du $k$-anonymat}

De multiples modèles ont été proposés afin de se prémunir contre les attaques
d'homogénéité, en particulier la
$\ell$-diversité~\cite{DBLP:journals/tkdd/MachanavajjhalaKGV07}
($\ell$-diversity), la $t$-proximité~\cite{DBLP:conf/icde/LiLV07}
($t$-closeness), la
$\delta$-divulgation~\cite{Brickell:2008:CPD:1401890.1401904}
($\delta$-disclosure), la
$\beta$-ressemblance~\cite{Cao:2012:PMR:2350229.2350255} ($\beta$-like\-ness).
Tous ces modèles rajoutent des contraintes sur les valeurs sensibles des classes
d'équivalence. Considérons la plus simple, la $\ell$-diversité.

\begin{definition}[$\ell$-diversité]
Une classe d'équivalence respecte la contrainte de $\ell$-diversité si elle
contient au moins $\ell$ valeurs ``représentatives'' pour la donnée sensible. Une base de données (ou
table) est dite $\ell$-diverse si toutes ses classes d'équivalence respectent la
contrainte de $\ell$-diversité.
\end{definition}

\begin{exemple}[$\ell$-diversité des footballeurs]
La Table~\ref{t4} est $2$-anonyme et $2$-diverse, car chaque classe
d'équivalence est composée d'au moins 2 n-uplets et chaque classe d'équivalence
est associée à au moins 2 valeurs sensibles différentes. A noter qu'il y a ici 3
classes d'équivalence : ([30;39], PSG), ([20;29], PSG) et ([32], OM).

La Table~\ref{t5} est certes $2$-anonyme, mais n'est que $1$-diverse puisque
pour la classe d'équivalence ([32], OM) il n'y a qu'une seule valeur sensible :
$500K$.
\end{exemple}

Il est ensuite possible de discuter de ce que signifie précisément
``représentatives'' ou combien de fois ces valeurs représentatives doivent
apparaître dans une classe d'équivalence pour que le critère de $\ell$-diversité
soit atteint. Si on comprend bien l'objectif que cherche à remplir ce modèle,
il faut prendre garde à traiter les données sensibles en prenant en compte leur
sémantique (d'où la question de données ``représentatives''). En effet, il ne
s'agit pas seulement d'avoir $\ell$ valeurs sensibles \emph{syntaxiquement}
différentes, encore faut-il qu'on ne puisse pas déduire des informations sensibles concernant les
utilisateurs, comme par exemple que tous les utilisateurs d'une classe
d'équivalence sont tous atteints d'une pathologie grave ou chronique.

\begin{definition}[$t$-proximité]
Une classe d'équivalence respecte la contrainte de $t$-proximité si la distance
entre la distribution de chaque attribut sensible de cette classe et la
distribution de chaque attribut sensible de la table complète ne dépasse pas un
seuil $t$. Une base de données (ou table) respecte la contrainte de
$t$-proximité si toutes ses classes d'équivalence respectent la contrainte de
$t$-proximité.
\end{definition}

\begin{exemple}[$t$-proximité]
Le développement d'un exemple de $t$-proximité étant un peu long, nous référons
le lecteur à l'article de Li \emph{et al.}~\cite{DBLP:conf/icde/LiLV07} pour un
exemple sur des données médicales.
\end{exemple}

La $t$-proximité précise la définition de ``représentativité'' des
valeurs, en obligeant la distribution des données sensibles de chaque classe
d'équivalence à ressembler, à un facteur $t$ près, à la distribution générale de
cette même donnée sensible. Commence à se poser alors la question de l'utilité
des données. Sous contrainte de $t$-proximité, les données ne paraissent pas
forcément directement exploitables. Toutefois, il reste possible de dégager des
tendances, ou effectuer des calculs généraux ou corrélations sur l'ensemble de
la table.

Il peut être délicat de savoir comment paramétrer la valeur du $t$ de ce modèle.
C'est l'objectif du modèle de $\delta$-divulgation : quantifier le gain
d’information d’un attaquant qui observe les classes d’équivalence, et qui
connait aussi la distribution des valeurs sensibles

\begin{definition}[$\delta$-divulgation]
Soit une valeur sensible $v_i$ avec une fréquence $p_i$ dans la base de données
originale, et une fréquence $q_{i,j}$ de cette valeur dans une classe
d’équivalence $Ec_j$.
La classe d’équivalence $Ec_j$ est dite $\delta$-disclosure-private ssi :
$\forall v_i, |\log(q_{i,j}/p_i)| < \delta$.
\end{definition}

\begin{exemple}[Les limites de la $\delta$-divulgation] Malheureusement, si
$p_i$ est grand et même pour un $\delta$ petit, il n’y a pas de borne maximale
réelle sur $q_i$ par exemple, on peut choisir $p_i = 0.5$ et $q_{i,j}=1$ et
$\delta = 0.5$ et on aura bien $\log (1/0.5) = log(2) = 0.3$, or si $q_{i,j}=1$
cela signifie qu'on connait avec certitude la valeur de la donnée sensible.
\end{exemple}

D'autres modèles existent permettant par exemple de mesurer certains critères
comme la probabilité d'appartenance d'un individu au jeu de données
échantillonné par rapport à un ensemble de personnes. C'est le cas de la
$\delta$-présence~\cite{Nergiz:2007:HPI:1247480.1247554} ($\delta$-presence). 

\subsection{Publications successives}

La publication successive d'un jeu de données (par exemple la liste des malades
d'un hôpital tous les mois) donne lieu à un problème difficile d'anonymisation.
En effet, la publication de deux jeux de données peut donner lieu à une attaque
par \emph{différence}, qui consiste a faire la différence entre les deux jeux de
données. Si l'attaquant connait les individus qui sont arrivés ou sont partis de
l'hôpital entretemps, il sera capable de remonter à l'ensemble de leurs données
sensibles. Certains modèles comme la
$m$-invariance~\cite{Xiao:2007:MTP:1247480.1247556} ont été proposés, mais leur
utilité est assez réduite, puisque les publications multiples accumulent des
données qui ont déjà été publiées, afin de se protéger. La meilleure manière de
faire des publications multiples sera d'utiliser la technique de confidentialité
différentielle que nous détaillons dans la suite.


\section{Differential Privacy}

\subsection{Intérêt et définition}
La DP\footnote{Nous utilisons ici le terme anglais, ou
son abréviation \emph{DP}. La traduction française communément admise est
\emph{confidentialité différentielle} }, introduite par Dwork~\cite{DBLP:conf/icalp/Dwork06,
DBLP:journals/fttcs/DworkR14} n'est pas un modèle d'anonymisation, mais la
caractéristique d'une opération (ou exécution d'un algorithme) sur des données
qui présente certaines garanties de confidentialité. Il est donc tout à fait
possible d'avoir des algorithmes cherchant à atteindre un modèle spécifique
d'anonymat, tout en proposant des garanties de DP. Par exemple, Domingo-Ferrer
et Soria-Comas montrent qu'il est possible d'atteindre des garanties de DP avec
le modèle de la $t$-closeness~\cite{DF15}.

La DP est très en vogue, car elle permet de quantifier un risque de
réidentification ``absolu''. Toutefois, le paramétrage de ce risque de
réidentification n'est pas simple, en particulier lorsque l'ensemble des
données sensibles est grand. En effet, la DP fonctionne au mieux lorsqu'il n'y a
qu'un faible nombre de valeurs pour les données sensibles (par exemple
vrai/faux, masculin/féminin, \ldots)

Considérons un algorithme $ALG$, et considérons deux bases de données
$\mathcal{D}_1$ et $\mathcal{D}_2$, telles que $\mathcal{D}_1 = \mathcal{D}_2
\cup d$ et $d \not\in D_2$. La garantie que cherche à fournir la DP est qu'en observant
$\Omega=ALG(\mathcal(D))$, le résultat de l'exécution de $ALG$ sur un jeu de
données $\mathcal{D}$, il sera \emph{très difficile} de savoir si
$\mathcal{D}=\mathcal{D}_1$ ou $\mathcal{D}=\mathcal{D}_2$.
Dit autrement, $\Omega$ ne doit pas changer beaucoup, selon que $d$ est présent
ou absent de la base de données utilisée en entrée de $ALG$. Un seul individu ne
doit donc pas foncièrement changer la valeur de l'exécution de $ALG$.

\begin{definition}[$\epsilon,\delta$-differential privacy]
Soient $\epsilon \in \mathbb{R}^{+}$ et $\delta\in\mathbb{R}^{*+}$.
On dit qu'un mécanisme $ALG$ respecte la contrainte de
$\epsilon,\delta$-differential privacy ssi $$Pr[ALG(\mathcal{D}_1)=\Omega] \leq
\exp(\epsilon)\times Pr[ALG(\mathcal{D}_2)=\Omega]+\delta$$ Si $\delta=0$ on
parle alors simplement d'$\epsilon$-differential privacy.
\end{definition}

Note : $Pr[ALG(\mathcal{D}_1)=\Omega]$ signifie ``la probabilité d'observer
$\Omega$ comme résultat de l'exécution de $ALG$ sur la base de données
$\mathcal{D}_1$.

Il découle de la définition de la DP que l'algorithme $ALG$ doit être un
algorithme \emph{probabiliste et non déterministe}, c'est-à-dire que plusieurs
exécutions successives de l'algorithme sur la même entrée peuvent produire des
résultats différents. En effet, si $ALG$ était déterministe, on ne pourrait pas
choisir $\epsilon$ aussi petit qu'on veut.

Il est également très important de souligner que les garanties de DP
s'ap\-pliquent également à tous les \emph{post-traitements}. C'est-à-dire que
tous les algorithmes d'analyse de données exécutés sur une base de données qui est
passée par un processus de $\epsilon$-differential privacy produiront des
résultats ayant cette même garantie, avec le même $\epsilon$.

\subsection{La DP en pratique}

La garantie proposée par le DP est que la probabilité d'observer une valeur
plutôt qu'une autre ne doit pas être sensiblement différente (i.e. à
$\exp(\epsilon)$ près) selon qu'un individu est présent ou pas. Nous allons voir
comment cela peut être appliqué en pratique. 

\begin{exemple}[Un exemple d'algorithme DP : la réponse aléatoire à un sondage]
Considérons l'algorithme classique de \emph{réponse aléatoire à un sondage} (ou
\emph{randomized response, RR}). Son objectif est de protéger les réponses des
individus à une question de type \texttt{vrai/faux} selon le processus suivant.

\begin{enumerate}
  \item La personne lance (secrêtement) une pièce de monnaie, si elle tombe sur
  face, elle donne la vraie réponse. (On considère ici une probabilité égale
  pile/face).
  \item Si la pièce de monnaie tombe sur pile, alors la personne relance une
  pièce. Si cette fois c'est face, elle répond \texttt{vrai}, si c'est pile,
  elle répond \texttt{faux}.
\end{enumerate}
\end{exemple}

Une rapide analyse montre que si une personne a répondu \texttt{vrai} à la
question, alors la probabilité que ce soit la vérité est de
$0.75$, alors que la probabilité qu'en vérité ce soit \texttt{faux} est de
$0.25$. Si ce qui nous intéresse est la proportion totale dans un ensemble de
personnes de \texttt{vrai/faux} alors on voit qu'on est en mesure d'estimer
cette valeur par rapport aux réponses observées : si $x$ est le nombre de
personnes ayant la propriété \texttt{vrai}, $x_O$ le nombre de personnes
observées ayant répondu \texttt{vrai} et $n$ le nombre total de personnes, alors
on est capable d'estimer $x \approx x_{est}=2\times x_O-n/2$.

Par rapport au $\epsilon$, il faut comparer les valeurs possibles d'espérance et
de variance pour les jeux de données contenant un individu particulier, et un
jeu de données ne le comprenant pas. Considérons l'objectif qui est de calculer
la fonction $x_{est}$, que l'on atteint en mesurant $x_O$. Toutes choses étant
égales par ailleurs, la réalisation de $x_O$ pour une base de données de
$\mathcal{D}_1$ contenant un individu ayant la caractéristique \texttt{vrai} et un jeu de données
$\mathcal{D}_2$ où il n'est pas présent est supérieure dans 75\% des cas et
identique dans 25\% des cas. Ainsi, avec les paramètres de notre exemple, on
aura 3 fois plus de chances de tomber juste en choisissant \texttt{vrai} si la
réalisation de l'algorithme est supérieure à l'espérance. En d'autres termes,
$\epsilon = \ln(3) \approx 1.09$.

Réciproquement, on peut aussi calculer la probabilité
$P$ qu'il faut appliquer au premier jet avant de relancer pour un $\epsilon$
donné. On pourrait montrer que pour des valeurs de $\epsilon$ très proches de 0,
il faut choisir $P=2\epsilon$.

\subsection{Evaluer la protection statistique à partir de $\epsilon$}

Il est intéressant de noter, par rapport à cet exemple, que nous avons discuté
de la valeur de $\epsilon$ mais que ce qui nous intéresse véritablement est le
risque de retrouver la valeur de la donnée sensible, soit ici la valeur
$P=0.75$. $P$ peut se calculer à partir de $\epsilon$, sachant que dans notre
exemple $P+\bar{P}=1$ et $P=exp(\epsilon)\times\bar{P}$, soit
$P=3\times(1-P)=0.75$.

D'une manière plus générale, il est également possible de remonter à cette
probabilité dans le cas où le nombre de valeurs de la donnée sensible est fini,
et vaut $n$.

\begin{theorem}[Probabilité de réidentification]
Soit $n$ valeurs pour une donnée sensible, et soit $x_O$ une observation d'un
algorithme respectant la contrainte de DP avec une valeur $\epsilon$. La
probabilité de pouvoir déduire la valeur de la donnée sensible à partir de cette
observation est de $P \leq \frac{exp(\epsilon)}{exp(\epsilon)+n-1}$
\end{theorem}
\begin{proof}
Soit $P_1$ la probabilité de retrouver la valeur correcte. Soient $P_2,\ldots,
P_n$ les autre probabilités. On a $\Sigma_{i\in[1;n]} P_i = 1$ et $\forall j>1,
P_1 \leq exp(\epsilon)\times P_j$. Si on considère que $\forall j>1, k>1, P_j =
P_k$ alors on a $P_1 \leq \frac{exp(\epsilon)}{exp(\epsilon)+n-1}$.
\end{proof}
On voit que dans le cas $n=2$ et $\epsilon = log(3)$ on retrouve bien $P_1 \leq
3/4$. On voit aussi, que pour des jeux de données où il y a un grand nombre de
valeurs de données sensibles, un $\epsilon$ de l'ordre de 10 peut fournir une
bonne protection. Par exemple si on a $n = 10^6$ et $\epsilon = 10$ alors
$P\approx 0.0216$.

\subsection{Théorème de composition}
La DP permet de résoudre le problème de la publication successive.

\begin{theorem}[Composition de la DP]
Un processus $ALG'(\mathcal{D})$ composé de l'exécution successive de $k$ fois
le même processus sur le même jeu de données $ALG(\mathcal{D})$ respectant la
contrainte d'$\epsilon$-DP respectera une contrainte de $(k\times\epsilon)$-DP.
\end{theorem}

Note : si les processus sont exécutés sur des jeux de données indépendants,
alors on prend simplement la valeur maximale des $\epsilon$ mis en jeu.

\begin{exemple}[Réponses aléatoires multiples à un sondage]
Reprenons l'exemple précédent, et appliquons exactement le même processus
aléatoire, mais deux fois de suite. Supposons que la réponse véritable soit
\texttt{vrai}. Nous répondons donc la vérité la première fois avec une
probabilité 0.75 et de même la seconde fois. La valeur de $\epsilon$ devient
donc $2\log(3)$. On comprend assez naturellement que $\epsilon$ augmente à
chaque fois qu'on observe de nouveau le résultat de l'algorithme, et en effet,
si on exécute l'algorithme un grand nombre de fois, il parait assez naturel de
penser que la valeur apparaissant le plus fréquemment sera la réponse véritable
de l'individu. 
\end{exemple}

Grâce à la DP, on est donc en mesure de quantifier très exactement le coût (en
termes d'augmentation du $\epsilon$ global) de la publication d'une donnée. 
Toutefois, le théorème de composition de la DP ne doit pas faire
croire qu'il est possible de publier à l'infini des données en respectant des
garanties fortes. Il est donc possible d'exploiter le théorème de deux
manières~:
\begin{enumerate}
  \item On se donne à l'avance un certain ``budget'' (sous la forme d'un
  $\epsilon$), et on s'autorise $k$ publications. Dans ce cas, chaque
  publication $p_i$ devra utiliser un $\epsilon_i = \epsilon / k$.
  \item On effectue $k$ publications avec divers $\epsilon_i$. Alors au final,
  on peut estimer le ``risque'' encouru $\epsilon = \Sigma_i \epsilon_i$.
\end{enumerate}

\subsection{Quelques exemples d'utilisation de la DP}

Suite à des attaques sur son recensement de 2000~\cite{DBLP:conf/pods/DinurN03},
le bureau du recensement américain a utilisé depuis 2018 des garanties de DP
pour son recensement de 2020~\cite{DBLP:conf/kdd/Abowd18}. L'un de leurs
problèmes principaux est de bien régler le compromis entre protection et utilité
des données.

En 2016, Google a proposé un framework nommé
RAPPOR~\cite{DBLP:journals/popets/FantiPE16} pour récupérer des données des
utilisateurs (afin d'effectuer des calculs de statistiques) avec des garanties
de DP. La valeur choisie dans ce cas de figure était de $\ln (3)$, avec une
approche inspirée de l'algorithme de \emph{réponse aléatoire}.

Apple a également été l'un des précurseurs à utiliser la DP dans ses algorithmes
d'apprentissage sur les données des utilisateurs des iPhones~\cite{apple}. Ils
ont appliqué leurs algorithmes d'IA au calcul des emojis les plus fréquents par
langue, aux profils de consommation énergétiques de leur navigateur, ou encore à
la découverte de nouveaux mots (comme des noms propres) pour le correcteur
orthographique. Dans ce cas de figure, les valeurs de $\epsilon$ utilisées sont
entre 2 et 8, pour une utilité d'apprentissage donnée.

Enfin, on pourra mettre en garde contre les approches qui cherchent à
maximiser l'utilité, sans toutefois utiliser la DP, mais qui peuvent
néanmoins être jugées conformes par rapport à la législation, comme
Diffix~\cite{DBLP:conf/apf/FrancisEM17}, et pour lesquelles des attaques de
désanonymisation ont été démontrées possibles~\cite{DBLP:conf/uss/GadottiHRLM19}.

\section{Logiciel de mise en \oe uvre des techniques d'ano\-nymisation}

Un outil \emph{open source} développé par l'Université Technologique de Munich,
ARX\footnote{ARX est disponible sur
:~\href{https://arx.deidentifier.org/}{https://arx.deidentifier.org/}}
\cite{DBLP:books/sp/15/PrasserK15} permet de réaliser des anonymisations selon de nombreux modèles à partir de
données originales en format tabulaire. L'outil permet également d'estimer les
risques de désanonymisation selon les modèles présentés plus haut, et même de
manière plus fine en calculant la distribution des probabilités de
désanonymisation et non simplement les valeurs maximales.



\section{Conclusion}

Cet article est introductif et ne décrit pas toutes les techniques
d'anonymisation qui existent.  Une description plus exhaustive peut etre trouvée
dans les articles suivants: \cite{Fung2010,ChenKLM09,Art29op}.

Nous avons essentiellement discuté dans cet article la question de
l'anonymisation appliquée aux données tabulaires. Il y a de nombreux autres
domaines, en particulier le domaine de l'anonymisation des données de
géolocalisation qui ont donné lieu à un grand nombre de publications ces
dernières années \cite{Nature13, MirICMW13bigdata,acs14,acs18}. Ces données sont
par nature difficiles à anonymiser car elles sont très identifiantes. En effet, des
études ont montré que la connaissance de 3 ou 4 points spatiaux-temporels d’une
trajectoire suffisait pour re-identifier, avec une probabilité élevée, une
personne dans une population de plusieurs millions d’individus\cite{Nature13}.
Different types d’anonymisation ont été proposé dans la litterature. Certaines solutions
proposent de publier uniquement des statistiques sur les differentes
trajectoires, comme leur longueur moyenne ou les endroits les plus souvent
visités. D’autres approches proposent de publier des donnees synthétiques,
c’est à dire des trajectoires  générées artificiellement à partir des
caractéristiques statistiques des vraies trajectoires\cite{DBLP:journals/tkde/AcsMCC19}.
Finalement, d’autres solutions proposent de modifier les trajectoires avant de les publier en, par
exemple, groupant les trajectoires similaires \cite{acs14} ou en y ajoutant du
bruit \cite{DBLP:conf/icde/AbulBN08}.

Lors du choix de la technique d'anonymisation à mettre en oeuvre, il convient de
rappeler que seule la DP permet d'avoir une garantie (probabiliste)
indépendante des connaissances des attaquants. Les autres modèles doivent faire
des hypothèses sur la connaissance des attaquants, et peuvent donc être parfois
contournés si un attaquant dispose d'informations qu'on ne l'imaginait pas
avoir (ou qui sont collectées postérieurement à l'anonymisation).

Une question difficile à trancher est la valeur du $\epsilon$ pour la DP. Nous
constatons que dans les faits, les entreprises mettant en oeuvre la DP optent
pour des valeurs autour de $\epsilon = 1$ ou plus, mais il faut bien comprendre
que la sécurité liée à cet $\epsilon$ dépend aussi du nombre de valeurs
différentes possibles parmi lesquelles il faut choisir. Ce n'est pas la même
chose d'avoir $\epsilon=1$ avec deux valeurs (auquel cas comme on l'a vu avec
l'Exemple 14 de la réponse aléatoire, une réponse pourra être très fréquente de
l'ordre de $75\%$ de chances d'observer la bonne valeur), et $\epsilon=1$ avec
1000 valeurs où toutes les probabilités auront une valeur de l'ordre de $10^{-1}$ avec une légèrement supérieure.
Quoi qu'il en soit, il faut choisir le plus petit $\epsilon$ possible permettant
de continuer à obtenir des résultats exploitables pour l'algorithme d'analyse de données.

Finalement, il faut rappeler que l'anonymisation ne constitue qu'un element dans une stratégie globale
de gouvernance des données qui doit aussi considérer, par exemple, l'audit des systèmes, 
la gestion du stockage des données ou la gestion des droits accès \cite{NAP18998}. 

\section*{Remerciements}
Les auteurs souhaitent remercier Christine Froideveaux pour sa suggestion de
l'écriture de cet article, pour sa relecture attentive et ses
commentaires, et Cédric Eichler pour des discussions autour de l'applicabilité
de la DP.

\bibliography{Bibliography}
\bibliographystyle{plain}

\end{document}